\documentclass[conference]{IEEEtran}%compsoc
\usepackage{amsmath,amsfonts}
\usepackage{algorithmic}
\usepackage{algorithm}
\usepackage{array}

\usepackage{textcomp}
\usepackage{url}
\usepackage{verbatim}
\usepackage{graphicx}
\usepackage{cite}
\usepackage{booktabs}
\usepackage{comment}
\usepackage{tcolorbox}

\usepackage{subcaption}

\usepackage{makecell}
\usepackage{multirow}
\usepackage{multicol}

\newcommand{\qw}[1]{\textcolor{blue}{#1}}

\usepackage{xcolor}
\usepackage{hyperref}

\usepackage{tabularx} % For tables with specified width

\begin{document}

%\title{CryptoCatch: Unmasking Encrypted\\ Mining Traffic}
%\title{A Two-Stage Encrypted Cryptomining Traffic Detection Mechanism}

\title{CryptoCatch: Cryptomining Hidden Nowhere}

\author{\IEEEauthorblockN{Ruisheng Shi\IEEEauthorrefmark{1}, Ziding Lin\IEEEauthorrefmark{1}, Haoran Sun\IEEEauthorrefmark{1}, Qin Wang\IEEEauthorrefmark{2}, Shihan Zhang, \\
Lina Lan\IEEEauthorrefmark{1}, Zhiyuan Peng\IEEEauthorrefmark{1}, Chenfeng Wang\IEEEauthorrefmark{1}
}
\IEEEauthorrefmark{1}\textit{Beijing University of Posts and Telecommunications} $|$
\IEEEauthorrefmark{2}\textit{UNSW Sydney} \\
}
\begin{comment}

\author{Ruisheng Shi, 
Ziding Lin,
Haoran Sun, 
Qin Wang, 
Shihan Zhang, \\
Lina Lan, 
Zhiyuan Peng, 
Chenfeng Wang,

\thanks{\textit{Ruisheng Shi, Ziding Lin, Haoran Sun, Shihan Zhang, Lina Lan, Zhiyuang Peng, Chenfeng Wang} are affliated with Beijing University of Posts and Telecommunications, China. 

\textit{Qin Wang} is affliated with UNSW Sydney, Australia.

Corresponding authors: \textit{Ruisheng Shi} (shiruisheng@bupt.edu.cn) and \textit{Lina Lan} (lanlina@bupt.edu.cn)}

}

\end{comment}

%\markboth{IEEE Transactions on Dependable and Secure Computing}%

\maketitle
\begin{abstract}
%Cryptomining behaviours pose severe security threats to society. However, existing blacklist and DPI-based techniques suffer from delayed blacklist updates and inability to identify encrypted cryptomining traffic. Furthermore, existing encrypted cryptomining traffic detection schemes usually fail to provide detailed information about cryptomining behaviours and do not have a solution to deal with false positives caused by detection models. In this paper, we propose a practical encrypted cryptomining traffic detection mechanism. It consists of a two-stage detection framework, which can effectively provide fine-grained detection results by machine learning and reduce false positives from classifiers through active probing. Unlike existing schemes, we perform active probing after the traffic classification to reduce false positives. Based on our collected dataset and extracted time series features, our classifiers detect mining traffic with an 0.99 F1-score and identify the cryptocurrency being mined with 99.39\% correct recognition rate. Furthermore, we have extensively evaluated the active probing scheme to verify its effectiveness for different mining pools. 

Cryptomining poses significant security risks, yet traditional detection methods like blacklists and Deep Packet Inspection (DPI) are often ineffective against encrypted mining traffic and suffer from high false positive rates. In this paper, we propose a practical encrypted cryptomining traffic detection mechanism. It consists of a two-stage detection framework, which can effectively provide fine-grained detection results by machine learning and reduce false positives from classifiers through active probing.  Our system achieves an F1-score of 0.99 and identifies specific cryptocurrencies with a 99.39\% accuracy rate. Extensive testing across various mining pools confirms the effectiveness of our approach, offering a more precise and reliable solution for identifying cryptomining activities.

\end{abstract}


%===================================
\section{Introduction} \label{sec1}

Dur Bitcoin's early development \cite{nakamoto2008bitcoin}, individuals could mine the cryptocurrencies using personal computing resources. This allowed for a decentralized entry point into the cryptocurrency world. However, the landscape of cryptomining has undergone a dramatic transformation due to the sector's swift growth, resulting in a significant increase in the network's cumulative hashrate. In November 2023, the Bitcoin network's total hashrate had escalated to an unprecedented 459 EH/s \cite{coinwarz2023}. This rapid increase made the prospects of profitable mining for individual or solo miners extremely slim. As a result, the vast majority of miners opted to join collaborative mining pools. The transition to pooling resources for mining activities was driven by the need to secure a reliable source of income. This enhances the likelihood of successfully validating new blocks and receiving mining rewards \cite{jablczynska2023energy}.

Additionally, while the mining industry has expanded, there has also been a noticeable increase in malicious activities by certain individuals. These activities involve embedding cryptomining scripts in websites \cite{papadopoulos2019truth} and installing malware on victims' devices \cite{hong2018you}. The attackers surreptitiously harness the victims' computing power for their own benefit, often incorporating these compromised devices into larger networks, or botnets, dedicated to mining cryptocurrency \cite{zareh2018botcointrap}. The illegal practices not only undermine the security of the digital landscape, but also pose significant risks to the broader goals of societal progress. These activities consume significant amounts of computing power and energy, running counter to global initiatives promoting energy efficiency and reducing carbon emissions. This presents obstacles to both technological progress and environmental sustainability.

There are two main risks in the field of cryptomining: insider misuse and external cyber-attacks \cite{cheng2022analysis}. Insiders might exploit equipment and electricity for active mining. Active mining can be categorized into two types: solo mining and pool mining. Solo mining involves individuals or small groups connecting directly to a blockchain's full node using specialized software \cite{xmrig2023}. Unlike solo mining, pool mining offers a collaborative platform for miners to combine their hashrate, enhancing the chances of solving blockchain puzzles. The mining pool allocates revenue based on each miner's hashrate contribution. When a blockchain puzzle is successfully solved, the pool distributes the rewards proportionally to the miners based on their individual contribution. 

Meanwhile, external attackers often target the devices in network, potentially commandeering them for cryptojacking operations \cite{meland2019experimental}. Cryptojacking is a malicious activity involving the unauthorized use of victims' devices for cryptomining \cite{eskandari2018first}. It manifests in two forms: browser-based cryptojacking and binary-based cryptojacking. Browser-based cryptojacking involved loading JavaScript or WebAssembly scripts in browsers, enabling mining by miners \cite{varlioglu2020cryptojacking}. Attackers embed mining scripts in websites, coercing visitors' devices into joining botnets for cryptomining. Binary-based cryptojacking is more covert. Binary malware clandestinely embeds itself in the victim's system \cite{hong2018you}, masquerading as legitimate background processes to mine continuously without detection \cite{tekiner2021sok}.

To safeguard their network, organizations often rely on simple and convenient methods, such as configuring firewall rules to block access to blacklisted URLs of mining pools and using Deep Packet Inspection (DPI) technology to detect and impede non-encrypted Stratum connections. However, this blacklist-based approach demands continuous maintenance and updates due to the ever-changing landscape of mining pool URLs. Maintaining an effective blacklist requires a stable and reliable source of threat intelligence, which might involve gathering public intelligence or deriving mining pool service URLs information from DPI detection results.

Several challenges arise from these methods. Firstly, the increasing use of proxy and private mining pools complicates the acquisition of blacklists \cite{zhang2023under}. Additionally, for security reasons, mining pools often change the URLs listed in blacklists for safety. Furthermore, DPI-based detection methods struggle to identify encrypted traffic, which poses a significant problem as most mining pools now offer SSL/TLS encryption services. Last but not least, supervisors often need more information about cryptomining behaviours to apply different governance measures. Therefore, we propose a robust detection solution that can detect encrypted cryptomining traffic and provide with more information related to mining behaviours.

\smallskip
\noindent\textbf{Contributions.}
To address the identified challenges in cryptomining traffic detection, we propose \textit{CryptoCatch}, a unified two-stage framework designed for detecting encrypted mining traffic in real-world networks.

Our main contributions are as follows:

\begin{itemize}
\item CryptoCatch combines flow-level time series classification with protocol-specific active probing, enabling high detection precision while reducing false positives. This tightly coupled design has rarely been explored in prior cryptomining detection work.

\item We design active handshake templates for commonly used mining protocols (e.g., Stratum-BTC, Stratum-ETH, Stratum-XMR), allowing real-time verification of suspicious mining pool destinations, which make a practical contribution for operational environments.

\item Our approach undergoes extensive evaluation. By selecting optimal time series features, our classifier achieves a 0.99 F1-score and accurately identifies the mined cryptocurrency with a 99.39\% accuracy rate. Beyond binary detection, CryptoCatch also supports multiclass classification to identify the specific mined cryptocurrency.

\item During the active probe stage, CryptoCatch showed a significant reduction in false positive rates, from industry averages of around 30\% to below 5\%, improving operational efficiency for network administrators and reducing unnecessary alerts. 
\end{itemize}

%$Compared to existing detection schemes, our mechanism can detect encrypted cryptomining traffic while combining active probing to effectively reduce the model's false positive rate, which is more practical in real-world scenarios.
%\smallskip
%\noindent\textbf{Practical use case.} We further apply our solution to practical seneriocs by decting cryptomining activities within campus networks. \qw{xxxxx berif intro the implemetation and results}

%The rest of the paper is organized as follows: Section \uppercase\expandafter{\romannumeral 2} describes the threat model and we will complete following experiments in that scenario. In section \uppercase\expandafter{\romannumeral 3}, we describe our methodology in detail, including our two-stage detection mechanism, as well as a fine-grained cryptomining traffic detection scheme based on machine learning. Section \uppercase\expandafter{\romannumeral 4} describes our active probing scheme and the evaluation of it. Section \uppercase\expandafter{\romannumeral 5} finalizes the paper with the discussion and conclusion.

%============================
\section{Technical Background}
\label{sec2}
%============================

\subsection{Understanding Cryptomining} \label{sec2.1}

\noindent{{\bf{Miners}}: Miners are participants in the blockchain network responsible for validating transactions, packaging blocks, and maintaining network operations. Miners configure and operate mining equipment, establish connections with blockchain nodes or mining pool servers, continuously receive computational tasks from their peers, and submit the results of successful computations to obtain mining rewards.}

\smallskip
\noindent{{\bf{Mining pool}}: A mining pool is an organization where multiple miners join forces, collectively contributing their hashrate to increase the success rate of mining and share rewards based on their contributions. As mining difficulty increases, individual miners may need to spend significant time and resources to successfully mine a single block. To improve their chances of earning rewards, miners can choose to join a mining pool. When any miner in the pool successfully mines a block, the rewards are distributed to all members of the pool according to their contribution levels.}

Mining activities can be divided into two categories: \textit{active} and \textit{passive} mining. Active mining~\cite{kiayias2016blockchain,islam2021blockchain} refers to miners actively using mining software to connect with a mining pool or independently mine blocks, with all rewards belonging to the miner. It includes solo mining and pool mining. Passive mining, also known as cryptojacking~\cite{tahir2019browsers}, occurs without the victim's consent. It can be browser-based (malicious scripts control a device when visiting compromised websites) or Trojan-based (malware infects a device, enabling mining in the background, with most profits going to the attacker).

Mining pools can be categorized into three types: public pools, proxy pools, and private pools. Public pools are open to all miners and distribute rewards proportionally according to the contributed hashrate. Proxy pools serve as intermediaries between miners and larger public pools, often used to aggregate traffic or conceal miner identities. Private pools are limited to specific users or organizations, allowing them to maintain full control over configurations and rewards.

\subsection{Active Mining} \label{sec2.2}

\noindent{\bf{Solo mining.}} It refers to individuals or small teams connecting to full blockchain nodes to participate directly in mining, making it one of the simplest mining methods. However, for mainstream cryptocurrencies like Bitcoin and Monero, the highly competitive, high-hashrate mining environment has made solo mining largely unprofitable for individual miners.

\smallskip
\noindent{\bf{Pool mining.}} With the rise of cryptocurrencies and increasing mining difficulty, individual miners face challenges such as dispersed computational power and unstable earnings. To address these challenges, mining pools emerged, making pool mining the most mainstream mining method today. A mining pool is a logical organization where miners from different physical locations join by registering and collectively share their computational power to solve blockchain hash problems more efficiently. By joining a pool, miners significantly increase their chances of discovering new blocks and receive rewards based on proportional allocation models.

\subsection{(Adversarial) Passive Mining} \label{sec2.3}

\noindent{\bf{Browser-based mining.}} Browser-based mining~\cite{sood2014empirical} refers to loading JavaScript scripts in web browsers. The method existed since the early days of Bitcoin \cite{quigley_bitcoin} and was widely used by crypto projects. JavaScript is a widely supported scripting language across mainstream browsers and makes browser-based mining convenient — users simply visit a webpage running specific scripts to initiate mining.

Malicious actors exploited this approach by deploying mining scripts on websites. Visitors would unknowingly mine cryptocurrency as the scripts ran in the background, a practice known as browser-based cryptojacking~\cite{tahir2019browsers}. This allows malicious website operators to earn mining profits without incurring any costs, at the expense of unsuspecting users.

\smallskip
\noindent{\bf{Trojan-based mining.}} Trojan-based mining~\cite{sood2014empirical} is another common technique used by attackers. Compared to browser mining, Trojan-based mining features more sophisticated attack methods, a broader reach, and longer operational duration. Attackers can distribute Trojan software to victims' devices through methods such as phishing emails, file disguises, system vulnerabilities, and brute-force attacks. Once executed, the Trojan software connects the victim's device to a mining pool in the background to perform mining tasks.

\begin{figure}[!t]
\centering
\includegraphics[width=3.3in]{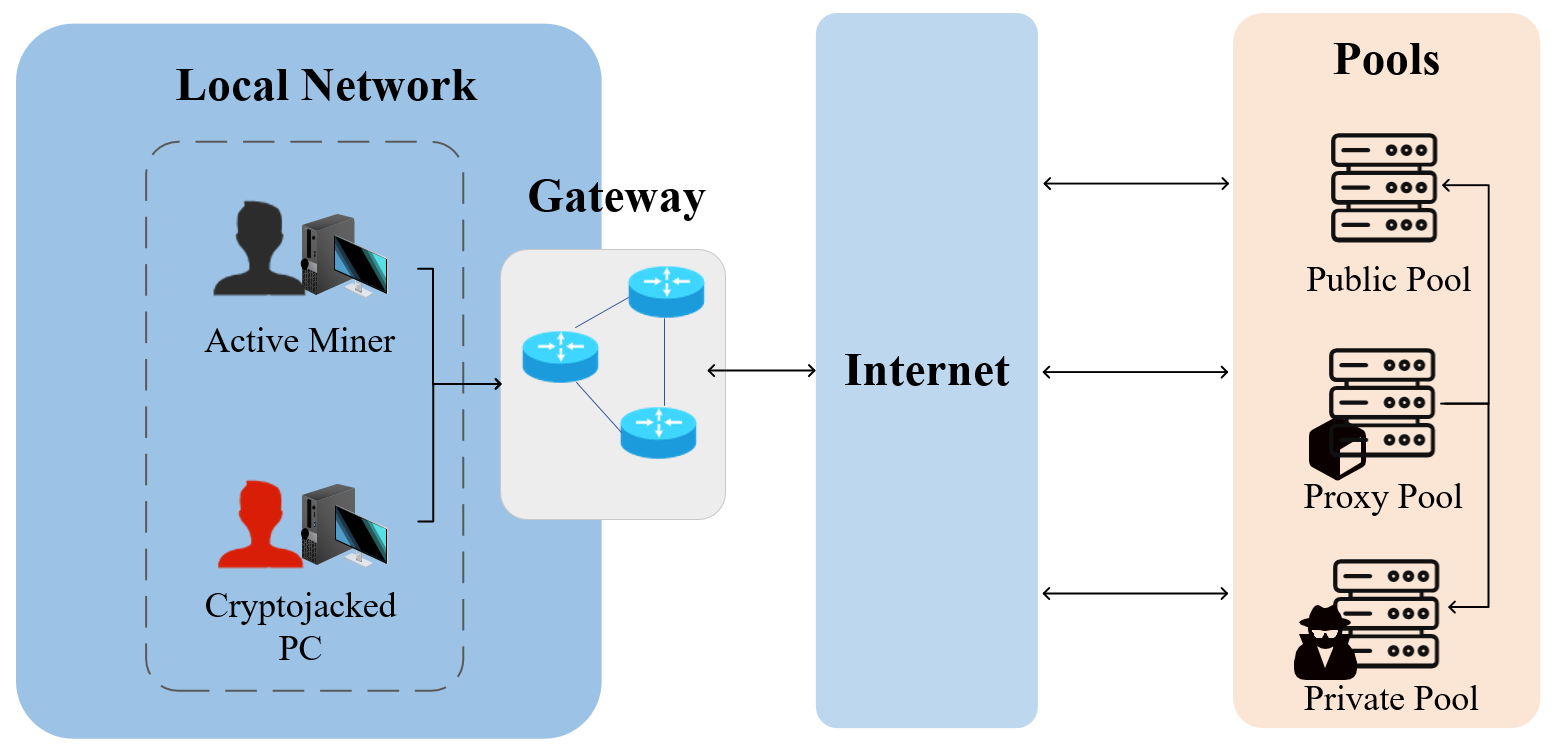}
\caption{Cryptomining activities in LAN}
\label{fig:threat_model}
\end{figure}

\subsection{Cryptomining Traffic Detection} \label{sec2.4}

Cryptomining detection methods are primarily based on \textit{host behavior}, \textit{browser} and \textit{machine learning} (ML).

Host behavior-based detection method~\cite{tanana2020behavior,gangwal2020detecting,xiao2023magtracer} try to spot cryptomining by looking for unusual host behaviours, e.g., CPU or memory usage. However, the latest mining software can configure a modest hashrate used for cryptomining by setting the throttle value. Also, these methods often need to install special detection tools on each device, which is expensive and impractical. 

Browser-based detection methods \cite{konoth2018minesweeper,kharraz2019outguard,naseem2021minos,saad2024analyzing} focus on different indicators, like WebAssembly (Wasm)~\cite{jangda2019not} or JavaScript (JS) file characteristics and their compilation times. These methods need access to Wasm, JS, and other files created during web browsing. They are good at finding cryptomining activities done through browsers, but they can't detect cryptomining that happens outside of browsers.

ML-based schemes \cite{tekiner2022lightweight,i2019detecting,pastor2020detection,caprolu2021cryptomining,hu2023detecting,Cao2023EarlyDetection,feng2022cj, Shi2024AnalysisPassiveActiveMoneroMining} are particularly relevant for most cryptomining scenarios, especially given the widespread use of SSL/TLS encryption by mining pools/software. Tekiner et al. \cite{tekiner2022lightweight} analyzed traffic from IoT devices running malicious mining software in smart home networks, applying classification models like SVM and GNB for cryptomining traffic detection. Mu{\~n}oz et al. \cite{i2019detecting} utilized NetFlow and IPFIX flow measurement techniques, using features of outgoing and incoming packet flow rates to differentiate cryptomining traffic from normal. Pastor et al.~\cite{pastor2020detection} used 51 features extracted from the Tstat tool to train a classification model and evaluate it in realistic scenarios. Hu et al.~\cite{hu2023detecting} proposed CMD-KST, by using the K-S test to compare the time intervals and packet size distributions of network traffic, which can efficiently detect encrypted mining traffic under proxy conditions with high precision. Shi et al. \cite{Shi2024AnalysisPassiveActiveMoneroMining} proposed a Monero mining behaviour detection method by analyzing the mining communication protocol to extract feature information.

\smallskip

In summary, while existing ML-based cryptomining detection approaches have demonstrated the effectiveness of time-series features and traffic-level classifiers, they primarily focus on passive binary detection, distinguishing mining traffic from non-mining traffic. These methods often lack mechanisms for active confirmation, fine-grained classification, or real-time adaptability.

In contrast, our work introduces a two-stage detection architecture that combines passive classification with protocol-aware active probing, enabling CryptoCatch to validate suspicious destinations and significantly reduce false positives. Furthermore, we maintain a dynamically updated blacklist based on real-time detection results and support multi-class identification of different cryptocurrencies. These design enhancements make CryptoCatch more suitable for real-world deployment scenarios, where encrypted, evasive, and multi-protocol mining behaviour must be handled with high accuracy and operational awareness.

%==================================
\section{Our Solution: CryptoCatch}
\label{sec3}
%==================================

In this section, we show how CryptoCatch works. CryptoCatch is a two-stage encrypted cryptomining traffic detection mechanism. We present the details below.

\subsection{Threat Model} \label{sec-model}

Mining activities are complex~(Fig.\ref{fig:threat_model}). On the one hand, users within these networks may take advantage of these devices and free electricity to mine for personal interests. On the other hand, attackers can exploit malware or malicious websites to take control of these devices for mining. For any local area networks, our threat model assumes:

\begin{itemize}
\item{each device connects to the external network via through internal gateways;}
\item{all miners focus on mining profitable cryptocurrencies;}
\item{all cryptomining traffic uses SSL/TLS encryption;}
\item{all mining pools that devices connect to are external.}
\end{itemize}

Miners are supposed to act either actively or passively (via cryptojacked devices): 

\begin{itemize}
    \item Active miners utilize organization's resources (e.g., PCs, servers) to engage with mining pools, excluding solo mining due to its limited profitability.

    \item Passive miners use cryptojacked devices that were compromised. They may perform browser-based and binary-based cryptojacking on victim devices. 
\end{itemize}

External mining pools fall into three categories \cite{zhang2023under}: public pools, proxy pools, and private pools. Both active miners and passive miners are connected to public or private mining pools, either directly or indirectly through proxy mining pools. 

\smallskip
Based on the above, the model can adequately describe the various mining activities under LANs. Our goal is to detect the encrypted cryptomining traffic generated by such activities in this scenario.

\begin{figure*}[!t]
\centering
\includegraphics[width=7in]{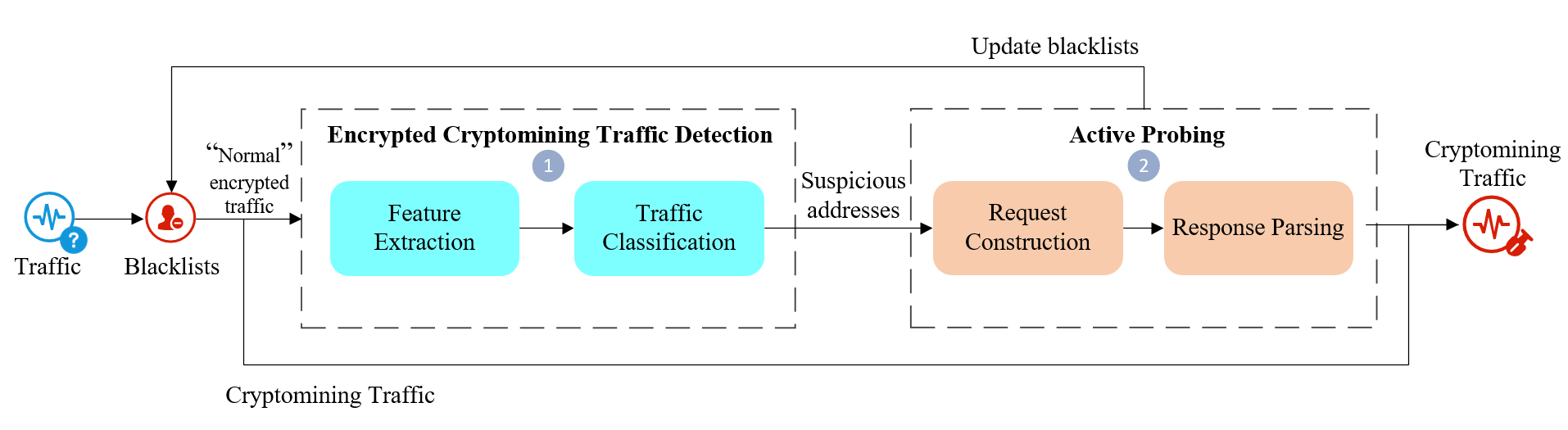}
\caption{The two-stage mechanism of encrypted cryptomining traffic detection}
\label{fig:two_stage}
\end{figure*}

\subsection{CryptoCatch in Two Stages} \label{sec4.1}

CryptoCatch operates in a two-stage detection pipeline (see Fig.\ref{fig:two_stage}). The first stage is a fine-grained ML-based traffic detection scheme, while the second stage is active probing.

\smallskip
\noindent\textbf{Stage1: Traffic detection and classification.} 
The first stage focuses on analyzing encrypted network traffic to detect mining activities with high precision. This stage adapts the FGTA methods \cite{feng2025unmasking} to the cryptomining domain, and leverages flow-level statistical and temporal features to classify whether a given network flow belongs to mining activity. The classification process consists of two key sub-modules.

The \textit{feature extraction} module collects a diverse set of statistical and time-series characteristics from encrypted traffic flows. Since cryptomining traffic is often persistent, periodic, and resource-intensive, we extract key features such as packet inter-arrival times, flow duration, byte distribution, and entropy metrics. These features help differentiate cryptomining traffic from general encrypted web traffic, which tends to be more diverse and less structured.

Once extracted, those features are fed into the \textit{traffic classification} module, which employs a trained machine learning classifier to categorize the traffic. The classifier not only detects cryptomining traffic but also identifies the specific cryptocurrency being mined based on protocol variations. The classification output provides a list of suspicious destination addresses and ports, representing the remote mining pools to which miners are connecting.
These extracted suspicious addresses serve as inputs for the second stage.

\smallskip
\noindent\textbf{Stage2: Active probing.} 
We employ active probing to validate whether the suspicious addresses extracted from the first stage genuinely belong to cryptomining pools. This process consists of two key sub-modules to confirm the nature of these addresses while minimizing false positives.

The \textit{request construction} module generates carefully crafted probing requests tailored to mining pool protocols. These requests mimic genuine mining client behaviors, ensuring that any mining servers in the target list respond as they would to legitimate miners. By simulating realistic interactions, this approach prevents evasive mining pools from easily detecting and blocking the probes.

The \textit{response parsing} module then analyzes the received responses for mining-specific patterns. This involves inspecting protocol headers, message formats, and response structures to determine whether the address operates as a mining pool. By comparing responses against known mining pool behaviors, we can differentiate mining pools from benign services.

In the Active probing stage, CryptoCatch maintains a dynamic blacklist of verified mining pool addresses, which is updated continuously based on classification and probing results. By cross-validating traffic classification results with the probing responses, we filter out false positives, ensuring that only confirmed mining pools are added to an updated blacklist. The system supports both real-time updates and daily batch updates depending on deployment needs. Real-time blacklist updates are suitable for environments requiring immediate detection, whereas batch updates are preferable during peak traffic periods or on resource-constrained gateways to minimize operational overhead. This update mechanism ensures the blacklist remains fresh and accurate, reducing the likelihood of stale entries or false positives. It is particularly effective in handling short-lived proxy pools or frequently rotating mining infrastructure, where traditional static blacklists fall short.

\subsection{Dataset Collection within Our Work} \label{sec4.2}
We focused on two datasets: \textit{normal traffic} and \textit{cryptomining traffic}. The latter is divided into active mining traffic and cryptojacking traffic. The former originates from direct connections to mining pools. Based on findings from \cite{eskandari2018first} and \cite{kharraz2019outguard}, which highlight Monero as a frequent target in cryptojacking, our cryptojacking traffic dataset centers on Monero mining via browser and binary malware.

The experiment was conducted in a controlled LAN network. Firstly, we deployed multiple virtual machines in bridged mode using VMware in each PC and server. Second, we deployed different mining software in each VM to perform cryptomining of different cryptocurrencies. Eventually, we run multiple VMs to perform cryptomining in parallel and efficiently capture the mining traffic at the gateway using Wireshark software. Due to the low volume of traffic generated by cryptomining, all mining activities lasted at least 6 hours. 

\begin{center} 
\tcbset{
        sharp corners=all, % Alternative to "enhanced"
        colback=gray!10!white,
        boxrule=0pt,
        colframe=gray!10!white,
        fonttitle=\bfseries
       }
       \begin{tcolorbox}
       \small
        \textbf{Responsible disclosure.} Before the mining process started, we had reported to the administration and obtained their permission to minimise the impact on the network. 
       \end{tcolorbox}
\end{center}

Eventually, we gathered cryptomining traffic for 7 different cryptocurrencies (Table~\ref{tab:dataset}). For active mining, we selected 7 categories of cryptocurrencies from the top 20~\cite{miningpoolstats2023}. These cryptocurrencies are popular among miners for their profitability. Note that although the Ethereum mainnet has transitioned to PoS after the Merge \cite{ethereum2022merge}, the Ethereum mining traffic in our dataset was collected from PoW-based Ethereum-compatible networks, specifically Ethereum Classic (ETC), EthereumPoW (ETHW), and EthereumFair (ETF), which continue to support active mining.
In contrast, our cryptojacking mining traffic dataset includes two main sources: browser-based and binary-based cryptojacking. We collect JavaScript code from existing browser-based mining services and incorporate it into our private websites. We then simulate browser-based cryptojacking behaviours by browsing these pages. For binary-based cryptojacking, we used public dataset from \cite{tekiner2022lightweight}, which used MinerGate to replicate malware intrusion scenarios. This dataset captures the entire traffic flow from the beginning to the end of cryptomining activities on victim devices. 

\begin{table}[!t]
\centering
\caption{Dataset of CryptoMining Traffic}
\label{tab:dataset}
\renewcommand{\arraystretch}{1.3} % Adjust line spacing
\resizebox{\linewidth}{!}{
\begin{tabular}{|c|cc|cc|}

\hline
\textbf{Type} & \textbf{Token} & \textbf{Provider} & \textbf{Time}/min & \textbf{Packet count} \\ 
\hline
\hline

\multirow{7}{*}{\makecell{Active\\ mining}} & BTC & F2pool & 296 & 288,440 \\
 & XMR & C3pool & 1,597 & 65,152 \\
 & ETC & F2pool & 3,174 & 55,296 \\
 & ETHW & JNpool & 2,361 & 84,736 \\
 & ETF & JNpool & 1,237 & 63,744 \\
 & CFX & JNpool & 838 & 50,240 \\
 & RVN & JNpool & 6,133 & 81,024 \\
\hline
\hline
\multirow{4}{*}{Cryptojacking} & XMR & \makecell{ Webminepool\\ (browser) }& 2,391 & 27,384 \\
 & XMR & \makecell{ Webmine \\ (browser)} & 1,879 & 27,874 \\
 & XMR & \makecell{ MinerGate\\ (binary)} & 984 & 22,111 \\
\hline
\multicolumn{2}{c}{} & \multicolumn{1}{c}{}  & \textbf{Total}  & \multicolumn{1}{c}{ 766,001 }\\
\end{tabular}
}
\end{table}

To simulate real scenarios, we use the part of CIC-IDS-2017 \cite{sharafaldin2018} as the normal traffic dataset, which contains benign behaviors of 25 users based on various protocols. We selected a representative subset of 803,334 packets from 2,263 traffic flows as our background data to create a balanced sample. 

For the collected dataset, we preprocess the dataset before extracting features, includes filtering the traffic data based on five-tuple and normalizing data. Since the main objective of this experiment is to detect the cryptomining traffic, we categorize the cryptomining traffic as positive samples and the normal traffic as negative samples. To ensure unbiased model training, we maintain a similar count of positive and negative samples. Lastly, for the experiment's subsequent phases, we split these samples into training and testing sets at a 4:1 ratio.

%============================
\section{Traffic Detection And Analysis}
\label{sec-experiments}
%============================

\subsection{Feature Extraction} 
\label{sec4.3}

SSL/TLS used in mining softwares and pools enhances communication security and impedes threat intelligence-based detection methods. However, distinct patterns emerge upon analyzing normal traffic and encrypted cryptomining traffic. As evidenced in \cite{vesely2019detect} and confirmed in our collected dataset, about 40\% of mining packets range from 36 to 80 bytes, and over 50\% fall between 105 to 110 bytes. This contrasts sharply with normal traffic, where less than 10\% of packets are between 105 and 110 bytes. Beyond packet length, cryptomining traffic differs significantly from normal traffic in terms of time-series characteristics, such as flow rate, peak transmission, and average packet length. Consequently, extracting time-series features from traffic is a viable and effective approach to differentiate normal traffic from cryptomining traffic.

We treat traffic packets as \textbf{time series} (Fig.\ref{fig:time}) that are continuously distributed over timestamps. We accordingly extract time-related features from the collected dataset based on three important pieces of information about traffic:

\begin{itemize}
\item{{\it{Timestamps}}: Timestamps indicate packet order and timing. Mining traffic often follows persistent high traffic patterns or periodic bursts, unlike the more irregular patterns of normal traffic.}
\item{{\it{Five-tuple}}: The five-tuple includes source and destination IP addresses, source and destination ports, and protocol type. Miners tend to communicate frequently with specific servers, which can be identified by their unique destination IPs and ports. Furthermore, different cryptomining activities may use distinct protocols, making protocol identification another useful classification factor.}
\item{{\it{Packet length}}: Cryptomining traffic typically involves intensive data exchanges, leading to recognizable packet length distributions. The temporal aspect of packet length variations can help identify traffic periodicity, burstiness, and continuity, which differ from normal network traffic.}
\end{itemize}

We first divide the traffic dataset into different flows based on the five-tuple. Each flow is then segmented into subgroups of 10 consecutive packets, denoted as $P = \{ pkt_{1},pkt_{2},\ldots,pkt_{10}\}$. If the last subgroup contains fewer than two packets, it is discarded, as individual packets do not provide meaningful time-series characteristics. {We adopt a 10-packet segmentation window following common practice in encrypted traffic classification work \cite{naseem2021minos}, which has been shown to offer a good trade-off between temporal granularity and statistical richness. Empirical evaluation shows that performance remains robust within a range of 8–12 packets; however, windows smaller than 5 packets lose periodicity cues, while larger windows (\textgreater 15) dilute burst characteristics unique to traffic.

For each packet in the subgroup, we extract the five-tuple, timestamp, and packet length as inputs to the feature extraction algorithm. This algorithm computes various time-series features, including basic statistical metrics (e.g., mean, standard deviation), time-window-based statistics (e.g., moving average, sliding standard deviation), spectral features from Fourier transforms (e.g., spectral coefficients, aggregated statistics), and autoregressive model features (e.g., autocorrelation, autoregressive coefficients). 

We use \textsf{tsfresh} \cite{blueyonder2023} to automate the feature extraction process, ensuring efficient analysis of those characteristics.

\begin{figure}[!t]
\centering
\includegraphics[width=3.3in]{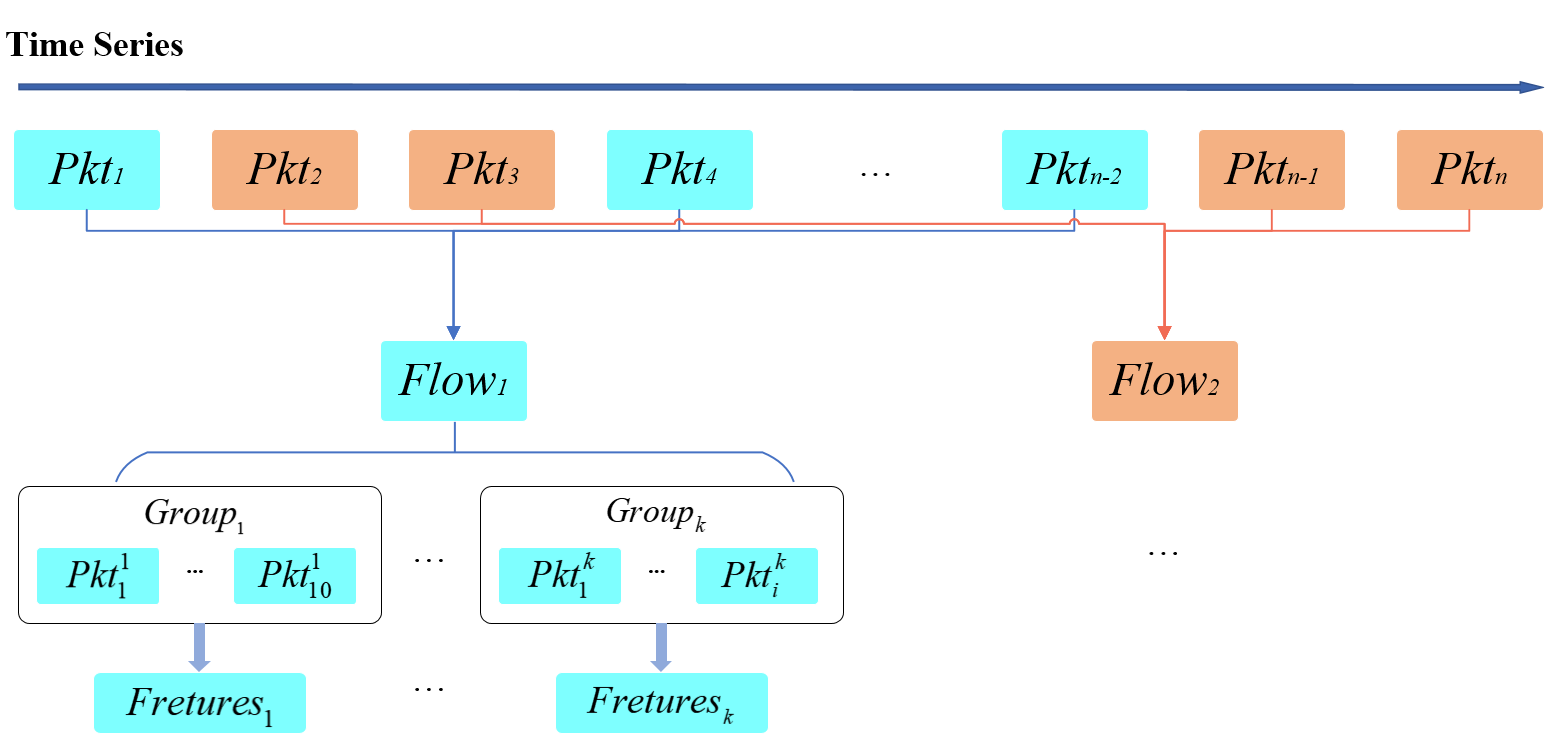}
\caption{Time series}
\label{fig:time}
\end{figure}

\subsection{Feature Selection} \label{sec4.4}

We then conducted the feature selection under different scenarios to further improve the model performance. 
The selected features are important to uncover the distinct characteristics of cryptomining traffic from various angles. These features aim to capture unique patterns inherent in mining activities. For instance, the total length of packets (sum-values) provides insights into the data-intensive nature of cryptomining traffic. The Number-cwt-peak, derived from Continuous Wavelet Transform (CWT). It also reflects the synchronization between computational and communication processes. The entropy value of packet length distribution (Binned-entropy-max-bins-10) and the quartile index quality (Index-mass-quantile-q-0.1) shed light on the regularity or concentration of packet length distribution in mining traffic.

We employed feature correlation analysis to identify the features that will be ultimately utilized for classification. The analysis is to assess the correlation between different features and the target variable.  In this paper, correlation analyses were conducted by combining univariate significance tests and the Benjamini-Hochberg method. The Benjamini–Hochberg (BH) method is a statistical procedure that adjusts p-value to control the false discovery rate when conducting multiple hypothesis tests. For each feature, a p-value was computed under the null hypothesis that the feature is independent of the traffic category label. Because testing many features increases false discovery rate, the BH method was used to adjust the p-values and control the false discovery rate. Features with p-values below 0.01 were considered statistically significant.

Before training the classifiers, we performed a correlation analysis on all the features and selected statistically significant features with a smaller p-value (less than 0.01). Finally, we selected 231 optimal features (partly shown in Table \ref{tab:feature}).

Since feature correlation analysis is done before model training, features with higher relevance scores do not fully determine their influence on classification. Therefore, we adopted a further feature importance analysis to analyze how much each feature contributes to the model prediction. This is often done during or after the model training and provides a basis for understanding the decision-making process of the model, feature selection, and improving model interpretability. 

\begin{table}[!]
\centering
\caption{(Selected) Optimal Features}
\renewcommand{\arraystretch}{1.3}
\begin{tabular}{|c|p{4.7cm}|}
\hline
\textbf{Feature Name} & \textbf{Description} \\
\hline
\hline
\multirow{2}{*}{Sum-values} & Total length of packets within the observation window. \\
\hline
\multirow{2}{*}{Number-cwt-peaks} & Number of different peaks in the packet length sequence. \\
\hline
\multirow{2}{*}{Binned-entropy-max-bins-10} & Entropy value of the packet length distribution divided into 10 bins. \\
\hline
\multirow{2}{*}{Index-mass-quantile-q-0.1} & The mass index corresponding to the 10\% quantile of the packet length distribution. \\
\hline
\end{tabular}
\label{tab:feature}
\end{table}

\subsection{Machine Learning Models} \label{sec4.5}
We focused on evaluating three advanced machine learning models: Gradient Boosting Machine (GBM), Random Forest (RF) and eXtreme Gradient Boosting (XGBoost) \cite{chen2016xgboost}. For all models, we utilized the default settings provided by scikit-learn or dedicated library for XGBoost. 

To evaluate the effectiveness of these models, we selected three evaluation metrics as below:

\begin{itemize}
    \item \textbf{Precision} quantifies the correctness of identified positives within the context of mining traffic predictions. It is the ratio of true positive (TP) outcomes (correct predictions of high mining traffic) to the total positive predictions made, both correct and incorrect.
    \item \textbf{Recall} measures the model’s ability to capture all relevant high mining traffic instances within the dataset. It is calculated as the ratio of TP (correctly predicted high traffic cases) to the sum of TP and false negatives (high traffic cases missed by the model).
    \item \textbf{F1-score} synthesizes both precision and recall, providing a single metric that balances the precision's accuracy against the comprehensiveness of recall in mining traffic predictions. It is particularly useful in scenarios where both false positives (FP) and false negatives (FN) carry significant consequences.
\end{itemize}

Based on the 231 optimal features obtained from the correlation analysis, Table~\ref{tab:accuracy_metrics} shows the calculated accuracy metrics. The results demonstrate that XGBoost performs better than other models across all metrics. XGBoost achieves a precision, recall, and F1 Score of 0.99, with an AUC nearing 0.998. In contrast, GBM and Random Forest exhibit relatively lower performance in this experimental context, highlighting significant advantages of XGBoost. Unlike Random Forest and GBM, which are also ensemble learning models, XGBoost benefits from a more intricate model structure. This complexity enables XGBoost to adapt more effectively to varying data characteristics, especially in scenarios involving incomplete or noisy data. Therefore, XGBoost's robustness and adaptability make it more suitable for cryptomining traffic detection.

\begin{table}[b]
\centering
\caption{Overall Accuracy Metrics Against Dataset}
\renewcommand{\arraystretch}{1.3}
\begin{tabular}{|c|ccc|c|}
\hline
\textbf{Model} &   \textbf{Precision} & \textbf{Recall} & \textbf{F1-Score} & \textbf{AUC} \\
\hline
\hline
GBM &  0.97 & 0.97 & 0.97 & 0.988 \\
RF & 0.99 & 0.98 & 0.99 & 0.998 \\
\hline
\textbf{XGBoost} &   \textbf{0.99} & \textbf{0.99} & \textbf{0.99} & \textbf{0.999} \\
\hline
\end{tabular}
\label{tab:accuracy_metrics}
\end{table}

During the model training process, XGBoost collects and analyzes statistical data from each tree, such as the number of splits and the gain from each split, to calculate feature importance. This metric, ranging from 0 to 1, reflects a feature's contribution to the model; a higher score indicates a greater impact. After calculating importance scores for 231 features, we sort them in descending order. We find that the importance scores of the features ranked after the tenth place are all lower than 0.03. The contribution of the top 10 key features to the model classification is significantly greater than that of the features ranked lower. The names and scores of the top 10 features are displayed in Fig.\ref{fig:topFeature}. We refer to these high-ranking features as key features.

The analysis reveals that the top 10 key features predominantly fall into three categories: general statistical attributes of the time series (e.g., \textit{Sum\_values} represents the sum of the time series values; \textit{Mean\_7\_absolute\_max\_number} is the average of the seven highest values; \textit{C3\_lag} measures non-linearity using C3 statistics), frequency domain features (e.g., \textit{Fft\_coefficients} for Fourier transform coefficients), and specific domain features (e.g., \textit{Friedrich\_coefficients} associated with dynamic Friedrich coefficients). The importance scores for features beyond the top ten fall below 0.03, with top features like \textit{Max\_langevin\_fixed\_point} scoring 0.9182 and \textit{Friedrich\_coefficients} at 0.7735, indicating that these top features are significantly more influential in model training compared to those ranked lower.

To validate these findings, based on datasets in Table \ref{tab:dataset}, we select the top 5, 10, 20 key features as well as all 231 features to train the XGBoost binary classifier. Each set is performed five times and the average value of each index is taken. The training results are shown in Fig.\ref{fig:feature metrics}. In addition, the effect of different number of features on the time spent in each stage was also recorded, and the results are presented in Fig.\ref{fig:feature performance}. Among them, the feature extraction time is the time for feature extraction on the same dataset, and the model training and model prediction time are the training and prediction time for each fold of validation in the five-fold cross-validation. From the results, it is known that compared to using all 231 features (Recall=0.99, F1-Score=0.993), using the top 10 key features can achieve similar performance (Recall=0.99, F1-Score=0.971), while reducing the feature extraction time by 64\% and the model training time by 80\%.

\begin{figure}[!]
\centering
\includegraphics[width=0.8\linewidth]{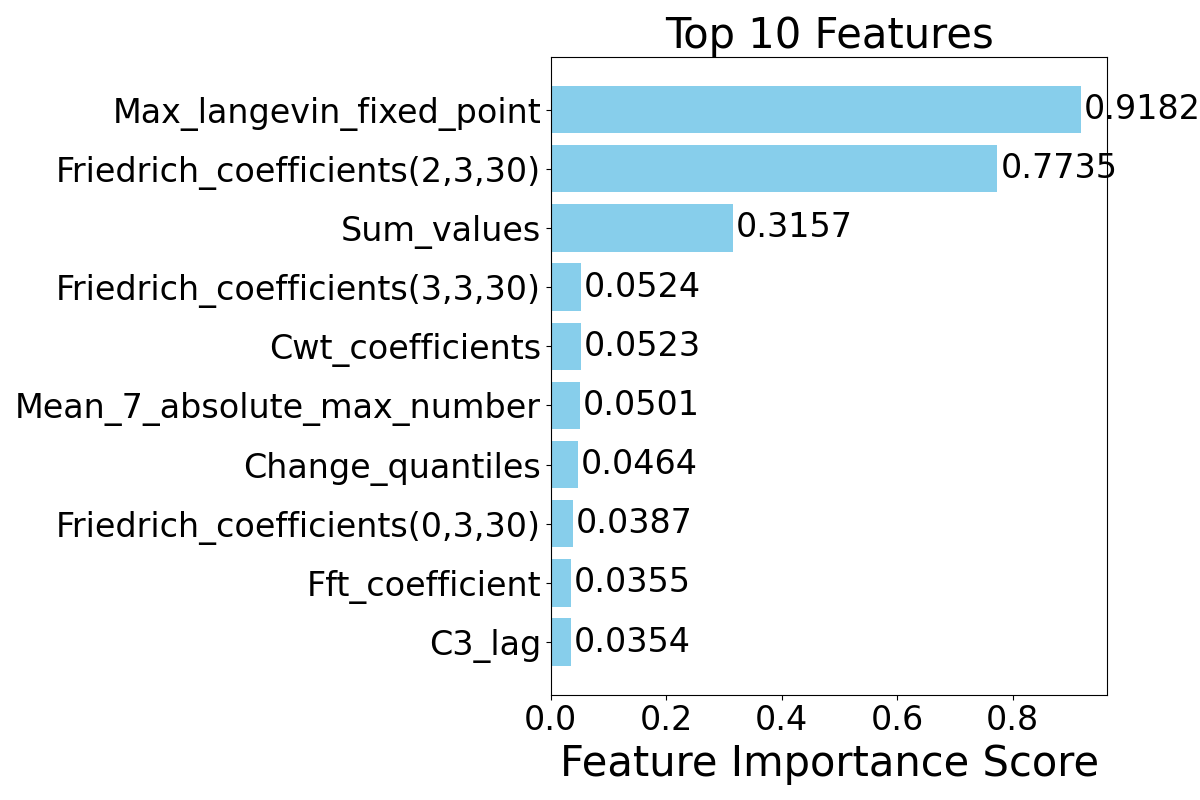}
\caption{Top-10 features and corresponding importance scores.}
\label{fig:topFeature}
\end{figure}

To obtain a highly sensitive classifier, we perform further experiments based on the effect of different classification thresholds on the metrics. The classification threshold takes values between 0 and 1. Based on the top 10 key features selected above, the changes of Precision, Recall and F1-Score are obtained by adjusting the classification threshold as shown in Fig.\ref{fig:threshold}. Meanwhile, based on the true positive rate and false positive rate calculated from different classification thresholds, the Receiver Operating Characteristic (ROC) curve is plotted as shown in Fig.\ref{fig:ROC}. From the experimental results, it can be seen that when the classification threshold is between 0 and 0.85, the sensitivity (recall) of XGBoost is maintained above 0.96 and decreases with the increase of the classification threshold. We further found that when the threshold is between 0 and 0.05, the precision and F1-score increase rapidly with the increase of the classification threshold, and increase to 0.93 and 0.96. When the threshold is between 0.05 and 0.85, the precision increases slowly, and the F1-score shows a tendency of first increasing and then decreasing. Meanwhile, in combination with the ROC curve, the XGBoost model with the classification threshold between 0 and 0.2 is able to achieve a high TP rate (0.99) with a low FP rate. Therefore, the XGBoost classifier with a classification threshold between 0.05 and 0.2 is able to achieve highly sensitive detection of mining traffic.

\begin{figure}[!t]
\centering
\begin{subfigure}[t]{0.7\linewidth}
\includegraphics[width=\linewidth]{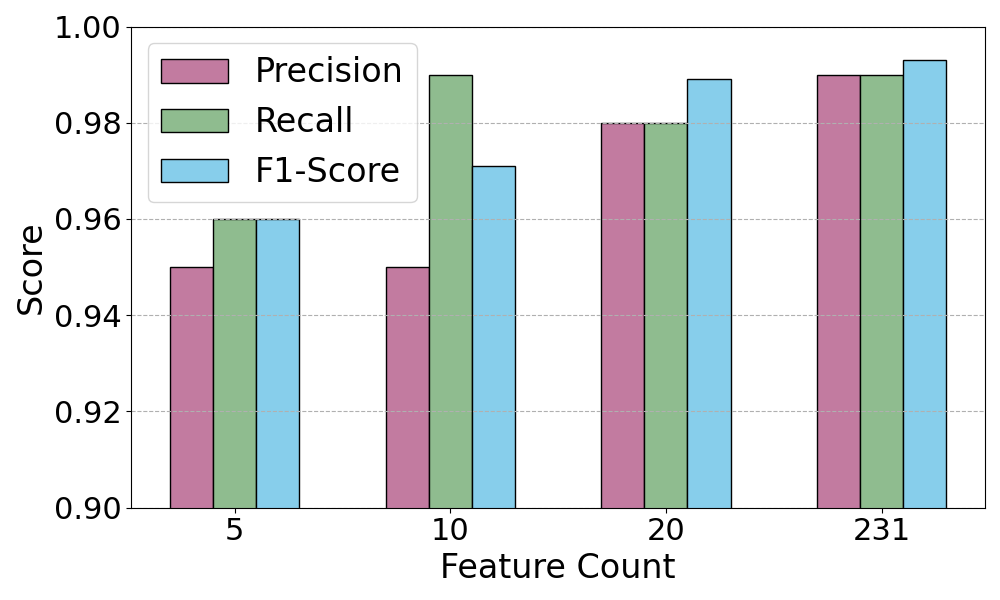}
\caption{Feature Metrics Performance}
\label{fig:feature metrics}
\end{subfigure}
\hfill
\begin{subfigure}[t]{0.7\linewidth}
\includegraphics[width=\linewidth]{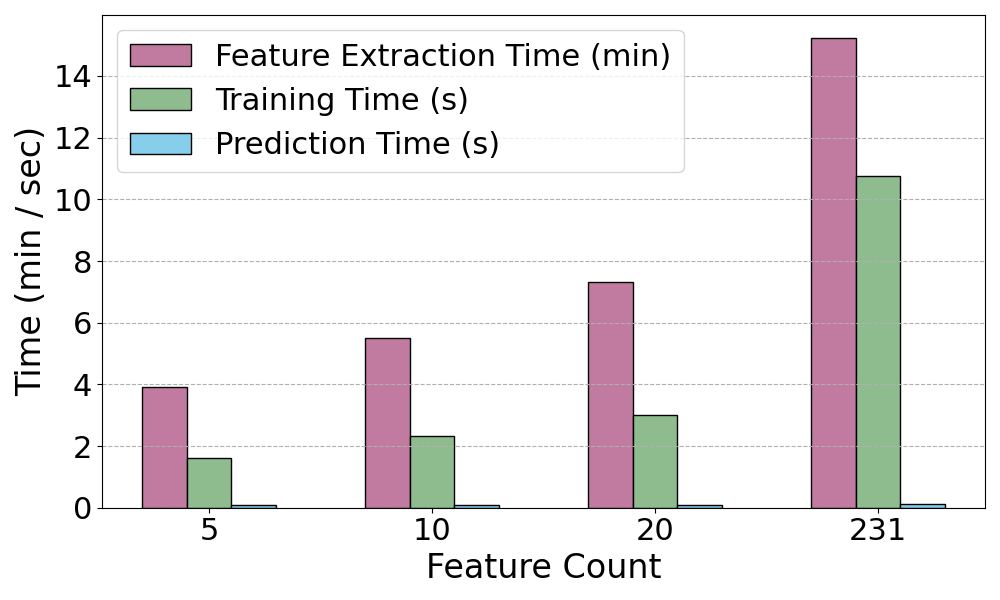}
\label{fig:feature training}
\caption{Time Comparison}
\label{fig:feature performance}
\end{subfigure}
    \caption{Feature Count Evaluation}
    \label{fig:var}
\end{figure}

%================================
\section{Active Probing} 
\label{sec5}
%================================

In Sec.\ref{sec3}, we demonstrate schemes to classify encrypted network traffic with machine learning models. These models effectively distinguish between normal and cryptomining traffic and identify specific cryptocurrencies involved in cryptomining. However, it already has a certain FP rate, where normal traffic is incorrectly classified as cryptomining behavior. To mitigate the issue, in this section, we propose an active probing scheme based on request construction and response parsing to detect mining pool services.

\begin{figure}[!t]
    \centering
    \begin{subfigure}[b]{0.75\linewidth} 
        \centering
        \includegraphics[width=\linewidth]{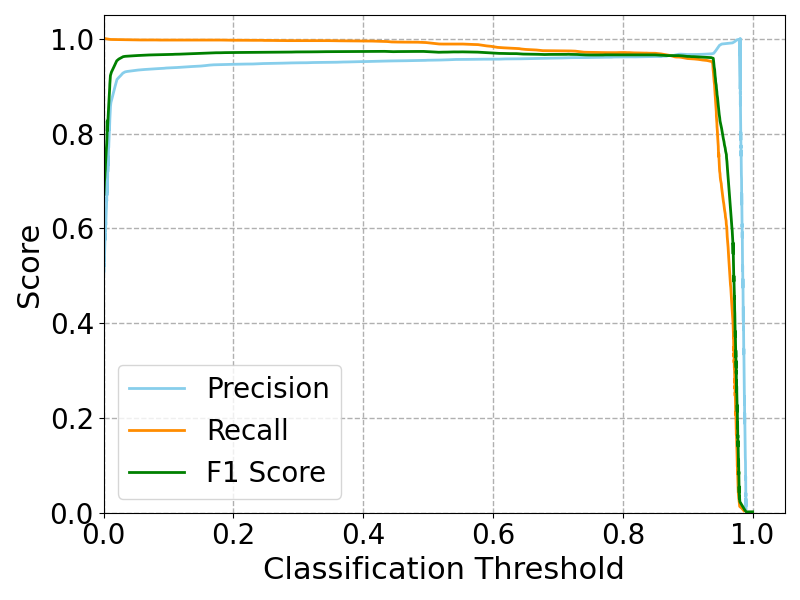}
        \caption{Threshold impact on PRF1 \label{fig:threshold}}
    \end{subfigure}
    \hfill
    \begin{subfigure}[b]{0.75\linewidth} 
        \centering
        \includegraphics[width=\linewidth]{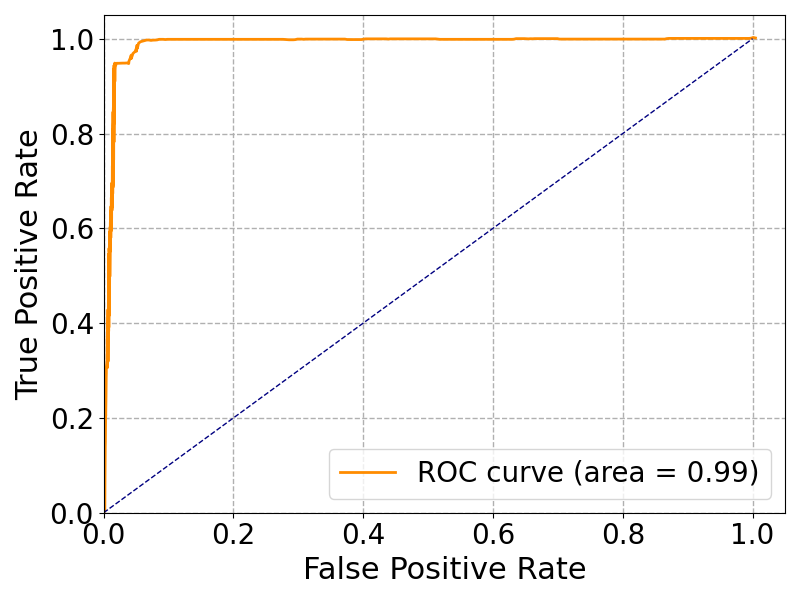}
        \caption{ROC curve \label{fig:ROC}}
    \end{subfigure}
    \caption{Comparison of evaluation metrics for different models}
    \label{fig:var}
\end{figure}

\subsection{Message Construction} \label{sec5.1}%Analysis of Mining Protocol
To construct suitable request messages, it is essential to understand the details of the communication between miners and mining pools. This involves grasping the message format and definitions as outlined by the mining protocol. Thus, we conducted a comprehensive analysis of mining protocols. Our approach is grounded in a thorough investigation of related papers (e.g., \cite{tekiner2021sok} and \cite{braiins2023}), source code from \cite{edsonayllon2020,sammy0072017,pooler2021}, and collected cryptomining traffic datasets. We found that the Stratum protocol \cite{braiins2023} is widely adopted in cryptomining for cryptocurrencies like Bitcoin, Monero, and Ethereum. The Stratum protocol is based on the TCP protocol and enhanced by JSON-RPC2.0. This combination addresses key communication challenges in the cryptomining sector, notably improving the interaction and efficiency of cryptomining.

The protocol is mainly structured around four message formats: miner subscription, miner authentication, mining job notification, and share submission \cite{braiins2023}. These formats facilitate a standardized collaboration process between mining pools and miners, allowing for the efficient distribution and completion of mining jobs. Our active probing method focuses on different specific implementations of the first three messages. Before cryptomining, miners usually need to send subscription and authentication messages to establish a connection with the mining pool. Then, the mining pool will return a response or mining job after possible authentication. The standardized communication is utilized in both active mining and cryptojacking scenarios. While the implementation of the Stratum protocol may vary slightly among different cryptocurrencies and mining pools, the fundamental pattern of the protocol remains consistent. %For example, Monero have adopted a refined version of the Stratum protocol, further optimizing the handshaking process for more efficient cryptomining operations (shown in Fig.\ref{fig:protocol}).

We focus on the Stratum protocols used by three major cryptocurrencies: Bitcoin, Monero, and Ethereum. Thus, we analyze their messages about miner subscription with each corresponding success and error responses, as illustrated in Table~\ref{tab:subsription}. By the way, there are two versions of the Stratum protocol, StratumV1 and StratumV2. However, StratumV2 is an unpractised protocol, it is only used by a few pools, such as \cite{braiinspool2023}. Consequently, most mining pools and miners continue to use StratumV1. Therefore, we only focus on StratumV1.

It is worth noting that our probing module is designed to be extensible, allowing new protocol templates to be easily integrated as emerging mining techniques are discovered. Based on this extensibility, we intentionally focus our analysis and protocol matching on widely adopted mining protocols, ensuring practical coverage of real-world scenarios while preserving the system's adaptability to future protocol evolution.

\begin{comment}
\begin{figure}[!t]
\centering
\includegraphics[width=3.3in]{img/Stratum-XMR.png}
\caption{Monero’s Stratum Protocol (ignore some unimportant fields).}
\label{fig:protocol}
\end{figure}
\end{comment}

\subsection{Active Probing Method} \label{sec5.2}

Our active probing method is based on subscription message construction and response parsing. For each target URL, we construct subscription messages for Bitcoin (Stratum-BTC), Monero (including Stratum-XMR and Stratum-Webmine-XMR), and Ethereum (Stratum-ETH) as shown in Table~\ref{tab:subsription}. 

Before sending the message, we establish a TCP connection with the URL to confirm its availability. If it is confirmed to be alive, we continue our probing method. Then, we send our constructed subscription messages by the following methods: 
\begin{itemize}
    \item  HTTP for plaintext transmission using JSON-RPC2.0; 
    \item HTTPS for cryptomining traffic via SSL/TLS protocol;
    \item WebSocket for communication in browser/server mode and commonly used for browser-based mining.
\end{itemize}

All processes adhere to TCP-based protocols, necessitating initial verification of a TCP connection with the target URL. We utilize HTTP and HTTPS to transmit messages for Stratum protocols related to BTC, XMR, and ETH. In contrast, WebSocket is exclusively utilized for Stratum-Webmine-XMR, reflecting its common use in browser-based mining activities, particularly with the cryptocurrency Monero.

Our analysis focused on specific fields in the success and error responses. We find that standard Stratum protocol responses typically include key fields (e.g., id, result, and error). Utilizing this pattern, we match fields within the responses to check for these common elements. The presence of these fields allows us to determine that the probed URL is offering a mining pool service.

\begin{table*}[!t]
\centering
\caption{Different subscription messages in the Stratum Protocol, each with its corresponding success and error responses} \label{tab:subsription}
\renewcommand{\arraystretch}{1.3}
\begin{tabular}{|c| p{3.8cm} | p{6cm} | p{3.8cm}|} 
\hline
\multicolumn{1}{|c|}{\textbf{Protocol}} & \multicolumn{1}{c|}{\textbf{Subscription message}} & \multicolumn{1}{c|}{\textbf{Success response}} & \multicolumn{1}{c|}{\textbf{Error response}} \\
\hline
\hline
\multirow{3}{*}{Stratum-BTC} & \{\textbf{id}: 1, \newline\textbf{method}:  mining.subscribe,  \newline\textbf{params}: []\} & \{\textbf{id}: 1, \newline \textbf{result}: [[mining.set\_difficulty,  $<$Target difficulty$>$], [mining.notify,  $<$Wallet address$>$]],  \newline\textbf{error}: null\} & \{\textbf{id}: 1,  \newline\textbf{result}: false,  \newline\textbf{error}: [20,  Not supported]\} \\
\hline
\multirow{3}{*}{Stratum-XMR} & \{\textbf{id}: 1,  \textbf{jsonrpc}:  2.0,  \newline\textbf{method}:  login, \newline\textbf{params}: \{\textbf{login}:  $<$Wallet address$>$,   \textbf{pass}:  x\}\} & \{\textbf{id}: 1,  \textbf{jsonrpc}:  2.0,  \newline\textbf{result}: \{\textbf{id}:  $<$Miner ID$>$,  \textbf{job}: \{algo:  rx/0,   \textbf{blob}:  $<$Blob$>$,  \textbf{target}:  $<$Target difficulty$>$\},  \textbf{status}:  OK\},  \newline\textbf{error}: null\} & \{\textbf{id}: 1,  \textbf{jsonrpc}:  2.0,  \newline\textbf{error}: \{\textbf{code}: -1,  \textbf{message}:  Invalid address\}\} \\
\hline
\multirow{3}{*}{Stratum-ETH} & \{\textbf{id}: 1,  \textbf{jsonrpc}:  2.0,  \newline\textbf{method}:  eth\_submitLogin,  \newline\textbf{params}: [$<$Wallet address$>$]\} & \{\textbf{id}: 1,  \textbf{jsonrpc}:  2.0,  \textbf{result}: true,  \textbf{error}: null\} & \{\textbf{id}: 1,  \textbf{jsonrpc}:  2.0,  \textbf{result}: null,  \textbf{error}: \{\textbf{code}: -1,  \textbf{message}:  Invalid login\}\} \\
\hline
\multirow{3}{*}{Stratum-Webmine-XMR} & \{\textbf{id}:  start,  \textbf{m}:  start,  \newline\textbf{p}: \{\textbf{token}:  $<$Miner token$>$\},  subscribe: 1\} & \{\textbf{r}: \{  \textbf{subscribed}: 0\},  \textbf{id}:  start\} & \{\textbf{e}:  noRights,  \textbf{id}:  start\} \\
\hline
\end{tabular}
\end{table*}

\subsection{Evaluation} \label{sec5.3}
To evaluate the effectiveness of our active probing method, we collect a list of mining pool service URLs from various types of pools, including \textit{public}, \textit{proxy} and \textit{private} pools. The dataset for these URLs is detailed as follows: 

\smallskip
\noindent{\bf{Public mining pools}}. These mining pools offer services to miners through publicly accessible service URLs. We leveraged data from \cite{miningpoolstats2023} as of November 2023 to identify the top 10 mining pools for Bitcoin, Monero, and Ethereum series (including Ethereum Classic and Ethereum PoW), based on their hashrates. Our analysis revealed that the combined hashrate of the top 10 pools accounts for over 92\% of the total network hashrate for each respective cryptocurrency. Subsequently, we registered with these pools and collected a diverse set of 124 URLs for probing, which varied based on the cryptocurrency, geographic service area, and the use of SSL/TLS encryption, among other factors.

\smallskip
\noindent{\bf{Proxy mining pools}}. We move to proxy mining pools. These pools offer traffic forwarding services for miners to address security issues linked to direct pool connections. We gathered proxy pools’ service URLs using two approaches: self-establishing proxy mining pools and extracting C\&C server URLs from available browser-based mining scripts. 

We acquired proxy mining pool software through various public sources. Initially, we identified that some public pools, previously recognised in our research, independently develop proxy pool software for miners. From these, we collected proxy software from three public pools: \cite{btccom2022}\cite{braiinspoolproxy2023}, and \cite{viabtc2021}. Subsequently, we explored public code repositories like GitHub \cite{github2023}, excluding projects with fewer than 50 stars or those exhibiting high similarity to others. This led us to gather seven representative mining pool proxy software solutions. Among these, \cite{evilgeniusdot2023}\cite{zaxblog2023}\cite{char1esorz2022}\cite{gominerproxybtc2022}, and \cite{nicococococ2022} support both BTC and ETC, while \cite{braiins2023} and \cite{deepwn2019} are specific to XMR. All software solutions support Stratum protocol encryption and traffic forwarding. We deployed these software solutions on our Virtual Private Servers (VPS), configuring different ports for those proxies supporting multiple cryptocurrency types to provide distinct proxy services for each type.

Furthermore, for browser-based mining, we collected publicly available JavaScript script samples that continue to offer browser-based mining services. Through this method, we extracted Command and Control (C\&C) server URLs, leading to the identification of three proxy mining pool service URLs from \cite{webminepool2023}\cite{webmine2023}\cite{browsermine2023}.

Eventually, we collected 16 URLs from 13 different proxy mining pools for active probing.

\smallskip
\noindent{\bf{Private mining pools}}. We collected about 1,200 code repositories associated with mining pool software from GitHub. We ignore those repositories with i) less than 100 stars; ii) high similarity to top-ranked software; iii) source code published by public mining pools; iv) no updates for over a year; and v) lack of support for cryptocurrencies in the Bitcoin, Monero, and Ethereum series. Ultimately, we identified and deployed \cite{sammy0072017pool}, \cite{jtgrassie2023}, and \cite{miningcore2023} that fully met our criteria. We opened different ports to support diverse mining services, including different cryptocurrencies and whether to use SSL/TLS. As a result, we established 9 unique private mining pool service URLs for probing.

\begin{figure}[!]
\centering
\includegraphics[width=0.8\linewidth]{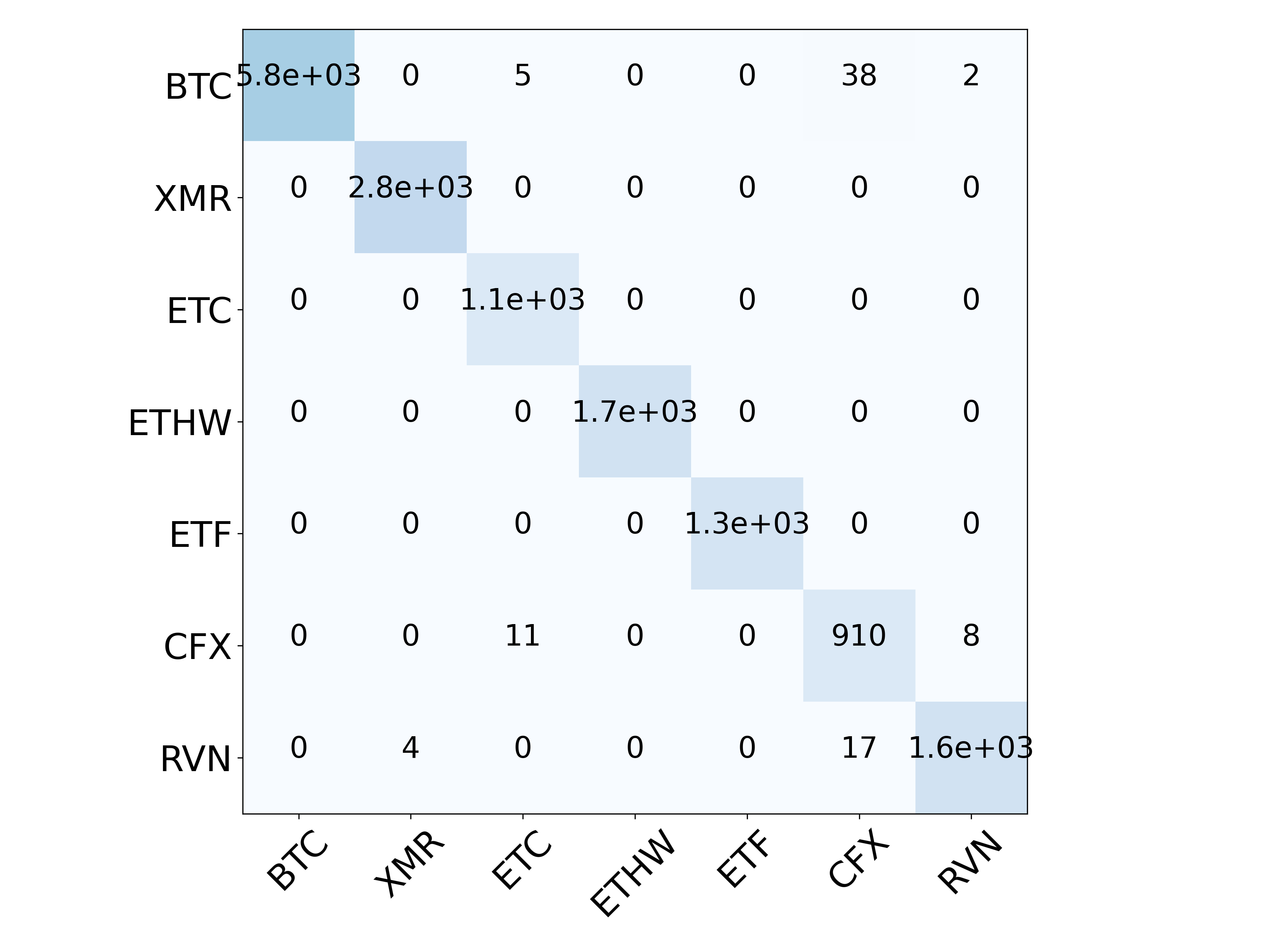}
\caption{Confusion matrix for the test set of the cryptocurrency detection experiment.}
\label{fig:matrix}
\end{figure}

\smallskip
\noindent{\bf{Result\&findings.}} We collected 149 mining pool service URLs. We probed this dataset using our active probing method based on request construction and response parsing. Ultimately, our probe successfully received success or error responses from 147 different mining pool service URLs. 

In addition, we also made new findings. First, the pools which were not successfully probed all belonged to browser-based mining pool services. It seems that these pools check the API token or site key information in the subscription message to verify the legitimacy of the request, and they will not respond to any messages with unregistered API tokens or site keys. This means that detecting browser-based mining pool services requires constructing request messages specific to different pools. In addition, for the public and private pools we collected, they do not take any response  measures to respond to incorrect requests but at least return an error response to inform miners. However, we have found from the source code of some private mining pools \cite{sammy0072017pool} that they can avoid this active probing by setting a blacklist or maximum number of connections. In this way, the pool does not respond to blacklisted miners and only sets a maximum number of connections that is just enough to reject unanticipated illegal probing requests. However, by default, the blacklist is empty, and the maximum number of connections is large enough for the private and proxy pools we deployed.

%=============================
\section{Extended Evaluation} 
\label{sec:extend}
%=============================-

\subsection{Deployment Overhead Evaluation} \label{sec deployment evaluation}

To evaluate the real-world applicability of CryptoCatch, we measured the system’s computational overhead, including feature extraction and active probing. All experiments were conducted on a commodity server equipped with an Intel Core i7-11700 CPU (2.50 GHz), 16 GB RAM, and Ubuntu 20.04.

\smallskip
\noindent{\bf{Active probing.}} The overhead of probing each suspicious destination address includes TCP connection setup and mining protocol-specific handshakes. Responsive mining pools typically respond within 75-120 milliseconds, while the worst-case timeout for unresponsive addresses is 255 milliseconds.

\smallskip
\noindent{\bf{Resource usage and throughput.}} CryptoCatch is capable of processing up to 4000 network flows per second. During sustained traffic conditions, CPU usage remains below 60\% and memory consumption does not exceed 8 GB.

In addition, the active probing component operates asynchronously and does not block flow classification, ensuring minimal impact on real-time detection. These findings suggest that CryptoCatch is efficient and lightweight enough for deployment on enterprise or campus gateway routers.

\subsection{Comparative Analysis} \label{comparative analysis}
Existing related works \cite{gangwal2020detecting}\cite{konoth2018minesweeper} usually choose the F1-score optimal classifier model to balance the precision and recall. Meanwhile, they usually do not make further determinations for the false alarms of the classifiers. Therefore, these schemes have poor practicability in real scenarios. In this section, we design comparison experiments with existing related work to demonstrate the effectiveness and practicality of the two-step mining traffic detection method proposed in this paper.

We employ the methodology presented in Sec.\ref{sec4.5} to train an XGBoost binary classifier. This classifier utilizes the top ten key features ranked by importance to classify network traffic. We adjust the classification threshold and observe the variations in the F1-score and recall of the classifier at different thresholds. Our objective is to identify the threshold that optimizes the F1-score or recall. 

Our selection strategies are as follows:

We determine the optimal classification threshold for the F1-score as the threshold that yields the highest average F1-score during five-fold cross-validation.
Notably, the highest sensitivity does not equate to optimal sensitivity. While the recall may increase with the reduction of the classification threshold, theoretically reaching a value of one by continuously adjusting, this occurs by disregarding other performance metrics such as the F1-score. The term  optimal sensitivity in this context signifies the need to find a balance where the recall is maximized while the F1-score remains relatively high and stable. To achieve this, we first establish a minimum acceptable value for the F1-score (set at 99\% of the optimal F1-score for this study) and seek the threshold that maximizes recall while maintaining an F1-score above this minimum. Secondly, to ensure that the F1-score remains stable as recall increases, the selection of the threshold must also consider the rate of change of the F1-score. In essence, the threshold is chosen to optimize recall without compromising the stability of the F1-score.

Following the selection strategy outlined above and in conjunction with the variation of evaluation metrics for the XGBoost classifier model at different classification thresholds as depicted in Fig.\ref{fig:threshold}, we draw the following conclusions: an optimal F1-score of 0.972 is achieved at a classification threshold of 0.42. When the threshold is reduced to 0.20, the F1-score meets the minimum acceptable value (approximately 0.962, calculated as 0.972×0.99) and remains stable, at which point the recall is optimized at 0.99. 

To compare the effectiveness of different approaches in detecting mining traffic, an additional dataset, as presented in Table~\ref{tab:analysis} was collected specifically for this comparative experiment. Within the same campus network environment as previous experiments, a total of 6 hours of traffic were captured at random intervals throughout a day, comprising 28,074 SSL/TLS encrypted data streams. This included 150 mining traffic data streams produced by active and passive mining of cryptocurrencies such as Monero and Ethereum, as well as 27,924 benign traffic data streams resulting from normal internet activities like web browsing, video conferencing, online videos, and file downloading. All mining pool IP addresses were known, enabling the isolation of all 150 mining traffic data streams.

\begin{table}[!]
  \centering
  \caption{Dataset for Comparative Analysis}\label{tab:analysis}
  \renewcommand{\arraystretch}{1.3}
  \resizebox{\linewidth}{!}{
  \begin{tabular}{|c|p{5cm}|c|}
\hline
    \textbf{Type of Traffic} & \textbf{Description} & \textbf{Count}\\
\hline
\hline
    SSL/TLS & Total encrypted data streams captured & 28,074 \\
\hline
    Mining  & Monero, Ethereum, etc. mining activities & 150 \\
\hline
   \multirow{2}{*}{ Benign }& Web browsing, video conferences, online videos, file downloading & 27,924 \\
\hline
  \end{tabular}
  }
\end{table}

Subsequently, four experimental scenarios were evaluated: a) optimal F1-score without active probe, b) optimal F1-score with active probe, c) optimal sensitivity without active probe, and d) optimal sensitivity with active probe. The active probe, to be detailed in Sec.\ref{sec5}, is designed to actively probe the destination IP addresses of traffic predicted as positive samples to confirm their status as actual mining pool addresses, thereby reducing false positives generated by the classifier.

Experiments conducted with the collected traffic data yielded confusion matrices for the four scenarios, as illustrated in Fig.\ref{fig:com}. The results indicate that Scenario c) had a higher false-positive rate (3.85\%) compared to Scenarios a) and b); however, by integrating active probing, all false positives could be resolved. Furthermore, Scenarios
c) and d) identified 132 mining traffic data streams, which is more than what was recognized by Scenarios a) and b). While Scenario a) had fewer false positives, it failed to detect 20 instances of mining traffic, which may not satisfy the requirements of supervisors in real monitoring scenarios. Moreover, since Scenarios c) and d) both incorporate the active detection module, Scenario d) effectively addresses the classifier-induced false positives, significantly enhancing the efficacy and practicality of this two-step detection approach.

\begin{figure}[!t]
\centering
\subfloat[\qw{}]{\includegraphics[width=0.5\linewidth]{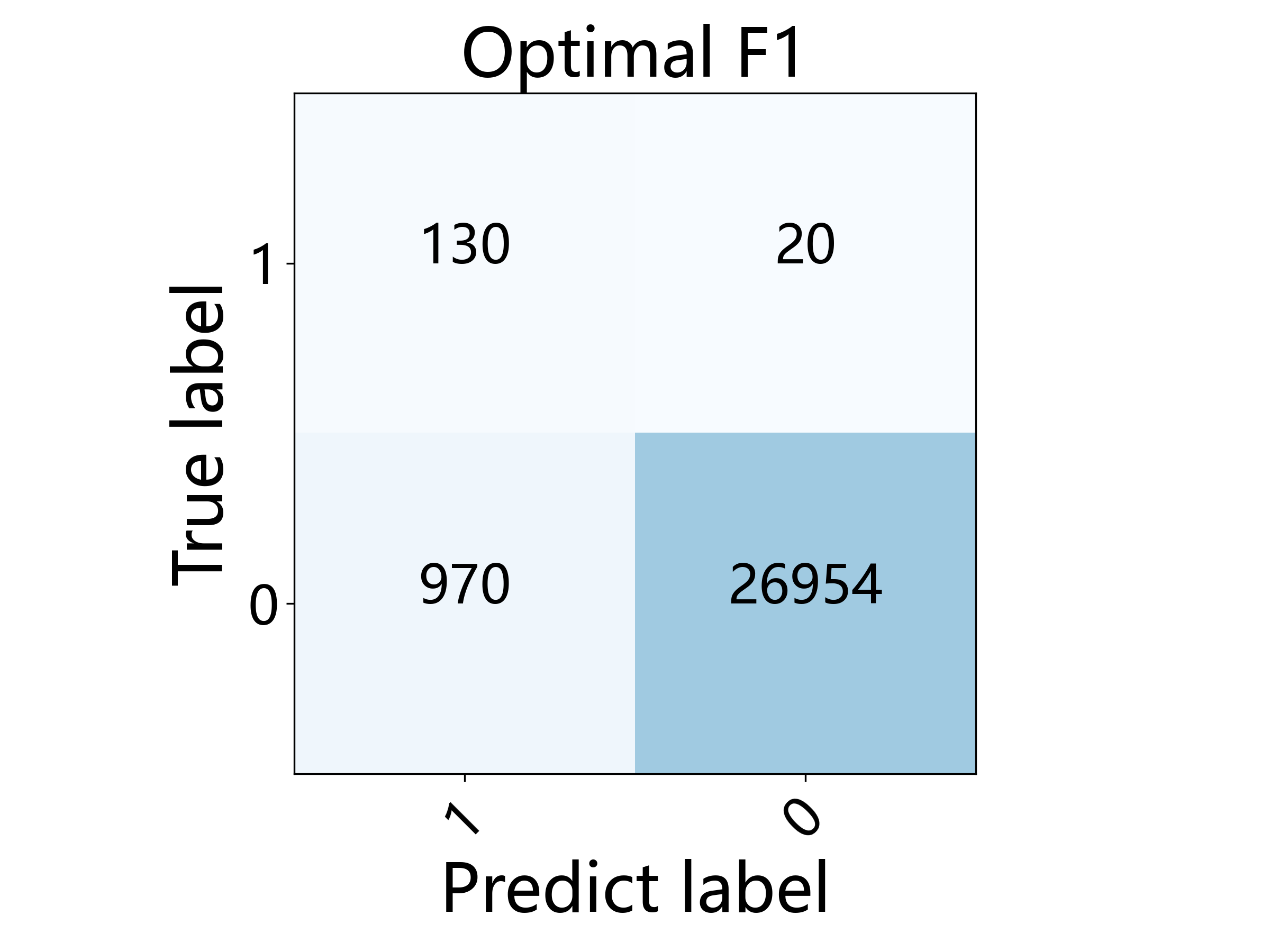}%
\label{fig:a}}
\hfill
\subfloat[\qw{}]{\includegraphics[width=0.5\linewidth]{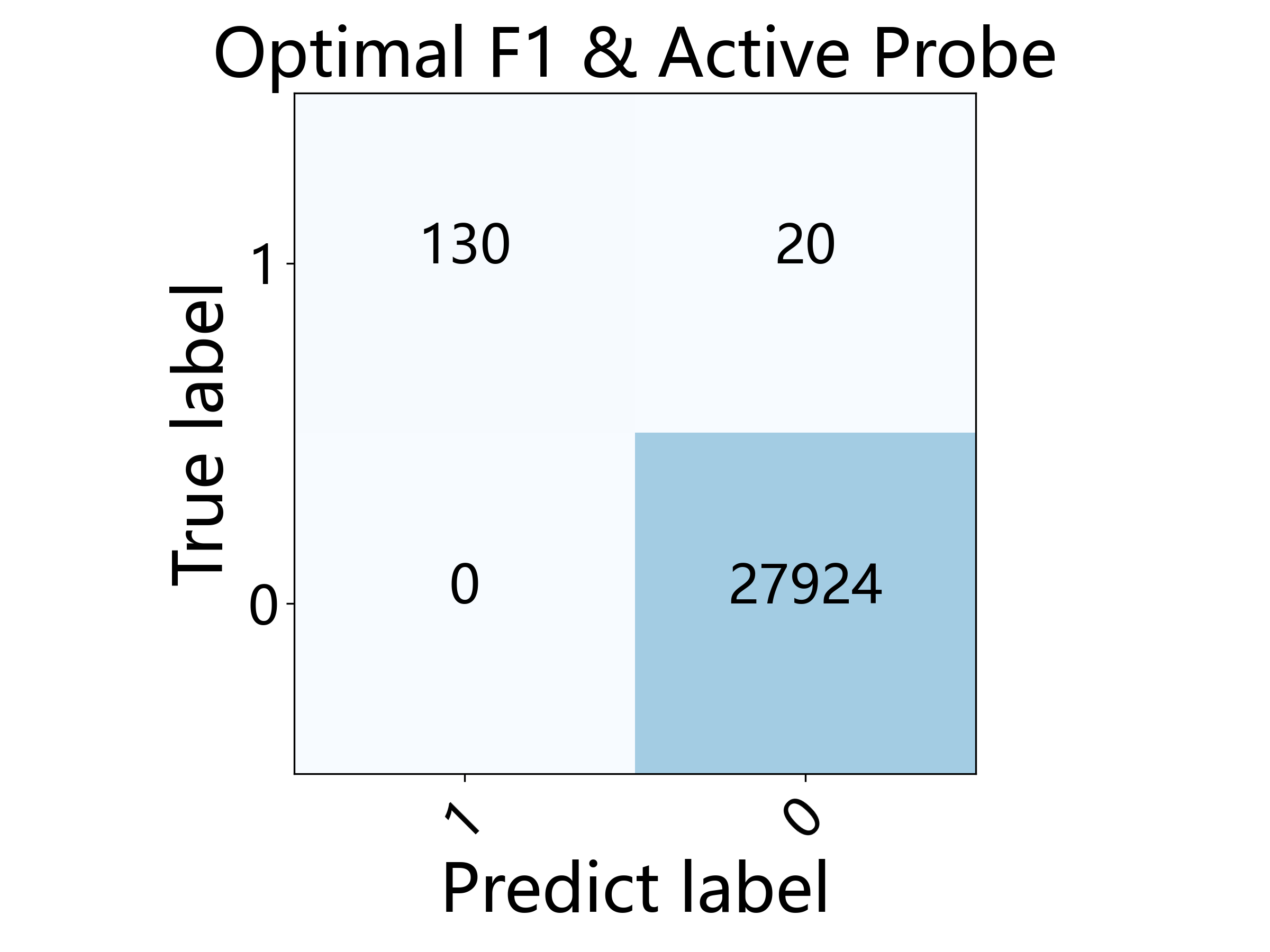}%
\label{fig:b}}
\hfill
\subfloat[\qw{}]{\includegraphics[width=0.5\linewidth]{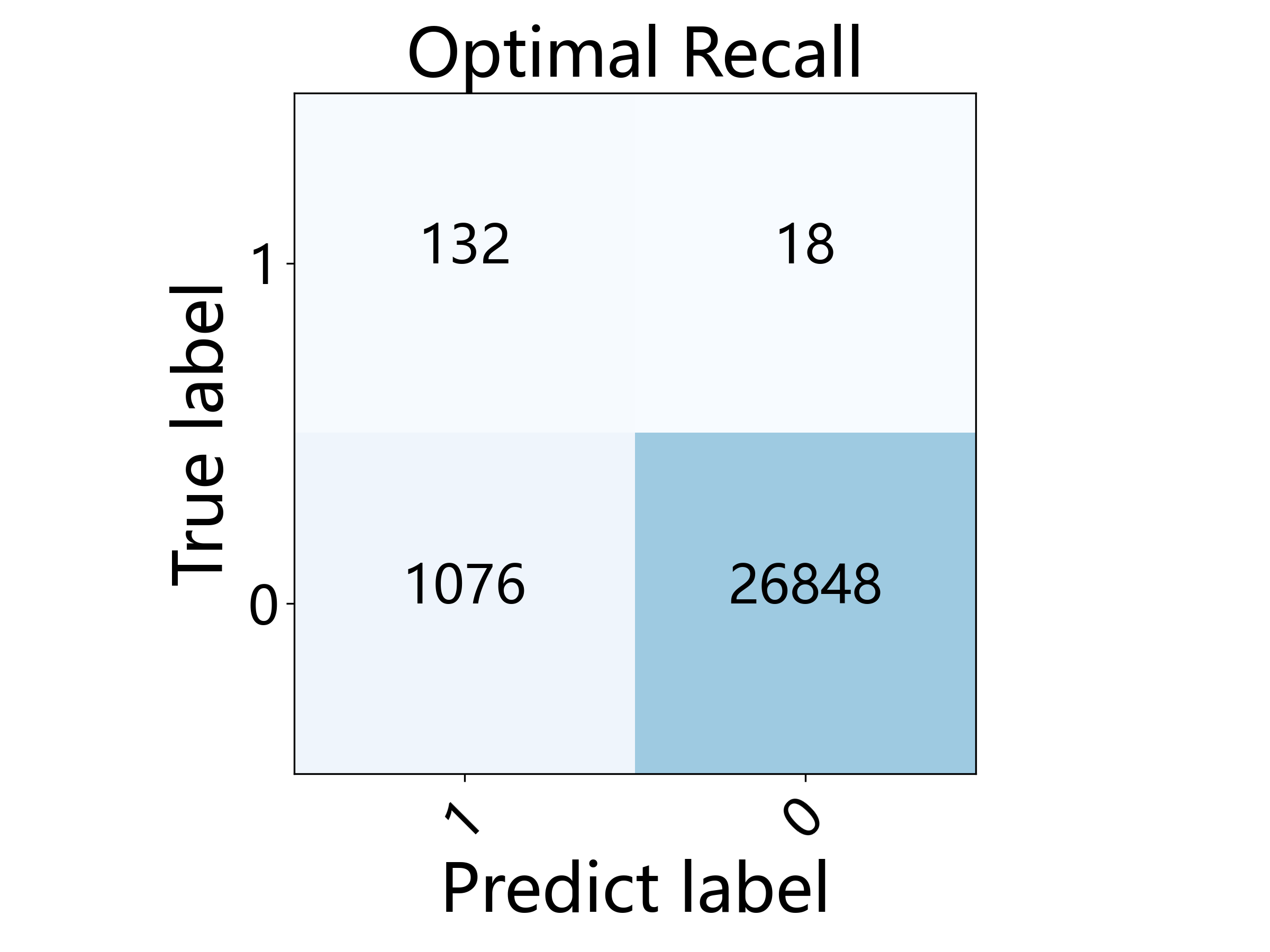}%
\label{fig:c}}
\hfill
\subfloat[\qw{}]{\includegraphics[width=0.5\linewidth]{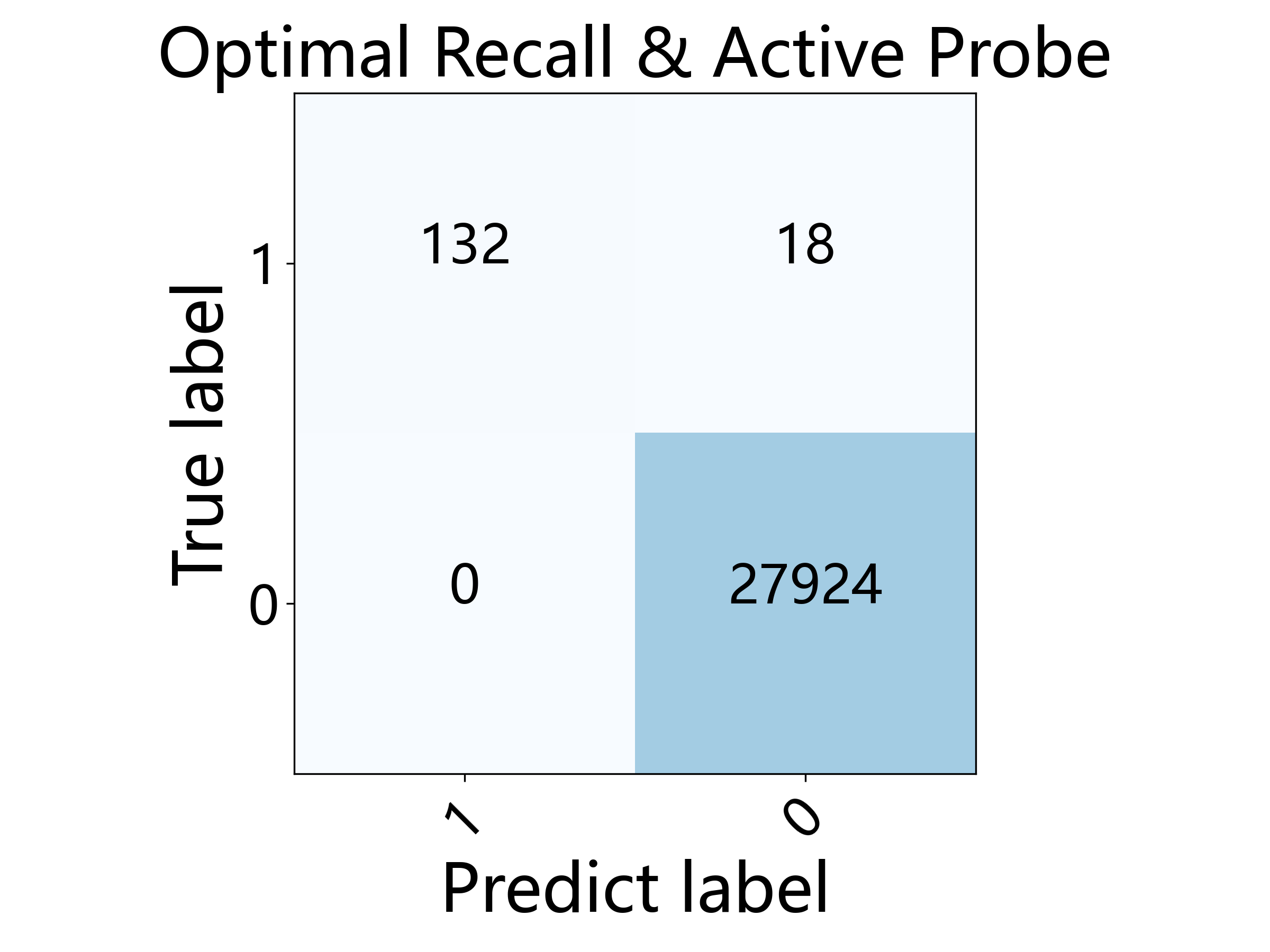}%
\label{fig:d}}

\caption{Confusion matrix for different detection scenarios.}
\label{fig:com}
\end{figure}

\subsection{Detection for Different Cryptocurrencies} \label{sec4.7}
Cryptocurrencies exhibit variations in their mining algorithms and protocols, which in turn influence their network traffic characteristics. For instance, Bitcoin utilizes the SHA-256 algorithm for mining, whereas Monero employs RandomX, enabling more accessible mining with ordinary hardware such as CPUs. We plan to delve deeper into these variations and their impacts on mining traffic (i.e., Sec.\ref{sec5}).

To accommodate these differences, our study incorporates cryptomining traffic data from seven distinct cryptocurrencies (Sec.\ref{sec4.1}). We have extracted 182 optimal time series features specific to cryptomining traffic, employing the feature extraction techniques (Sec.\ref{sec4.2}). These features are then utilized to train an XGBoost model, which is tasked with the multi-class classification of cryptomining traffic to identify specific cryptocurrencies.

The efficacy of the XGBoost model is heavily reliant on the precise tuning of its hyperparameters. To address this, we implement Bayesian optimization, an approach that surpasses conventional methods like random or grid search. Bayesian optimization efficiently narrows down the hyperparameter space and leverages historical data to form a Gaussian process model, significantly enhancing our ability to locate the optimal settings with fewer iterations.

We use softmax as the loss function for XGBoost. To validate the model's effectiveness and mitigate the risk of overfitting, we also implement five-fold cross-validation during the training and validation stages. This approach allows us to calculate multinomial log loss (mlogloss), as in:

\begin{equation}
\label{deqn_ex1a}
mlogloss = -\frac{1}{N} \sum_{i \in N} \sum_{j \in M} y_{ij} \log(p_{ij})
\end{equation}

In Equation~\ref{deqn_ex1a}, $N$ represents the total number of samples in the dataset, and $M$ indicates the number of categories (7 for this study). The term $y_{ij}$ is a binary indicator, assigned 1 if sample $i$ belongs to category $j$, and 0 otherwise. The term $p_{ij}$ represents the model's predicted probability that the sample $i$ belongs to the category$j$. We continuously adjust the range of hyperparameters and the number of Bayesian optimization iterations. Through repeated training and multiple experimental sets, we select the optimal results for model construction. 

We carried out 10 separate experiments on the dataset and calculated the average of the individual hyperparameters obtained. The gamma value was set to 0.01 to control split decisions and avoid unnecessary complexity. The maximum tree depth was set to 4, preventing overfitting. The subsample ratio of 0.819 ensured that each boosting round used a portion of the training data to improve generalisation. The column sample by tree was 0.514, limiting the number of features considered in each split.

\begin{figure}[!b]
\centering
\includegraphics[width=0.8\linewidth]{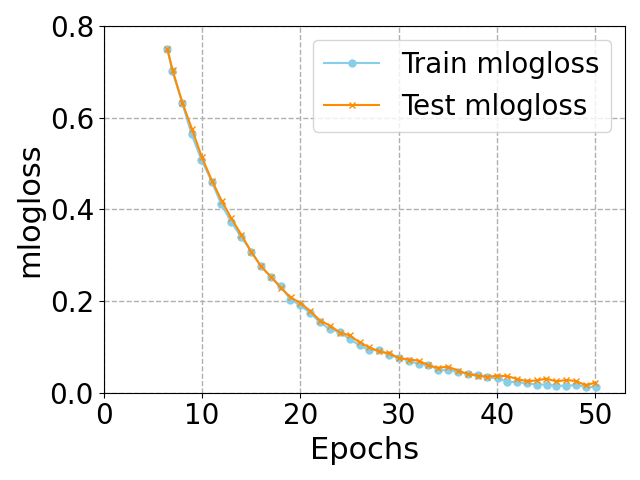}
\caption{Variation of mlogloss of XGBoost model}
\label{fig:mlogloss}
\end{figure}

For stability, we set the minimum child weight to 5, ensuring that each node had enough samples before splitting. The learning rate was set to 0.409, balancing training speed and accuracy. These settings helped the model achieve strong performance while keeping computation efficient.

The best parameter combination is presented as: 

\begin{center}

\fbox{%
    \begin{minipage}{0.9\linewidth}
\begin{itemize}
    \item \textbf{Gamma: 0.01}, which helps control overfitting by specifying a minimum loss reduction required to make a further partition on a leaf node.
    \item \textbf{Max depth: 4}, restricting the complexity and the depth of trees.
    \item \textbf{Subsample: 0.819}, where about 82\% of the training data is used for building trees, helping to prevent overfitting.
    \item \textbf{Colsample by tree: 0.514}, where 51.4\% features are used per tree, enhancing model performance by introducing randomness into feature selection.
    \item \textbf{Min child weight: 5}, defining the minimum sum of instance weight needed in a child to continue creating further partitions.
    \item \textbf{Learning rate: 0.409}, controlling the contribution of each tree, making the training process  robust by preventing rapid convergence to local minimum.
\end{itemize}
    \end{minipage}
}
\end{center}

Using the optimized hyperparameter settings, Fig.\ref{fig:mlogloss} shows the variation in the XGBoost model's mlogloss over the number of training epochs for both the training and testing datasets. Notably, the model achieves a low mlogloss of 0.079 for the training dataset and 0.083 for the testing dataset after 30 epochs, demonstrating its robust ability to fit the data well and generalize effectively across both datasets. Furthermore, the model attains a high correct recognition rate of 99.39\% after 50 epochs, evidencing its precision and effectiveness in classifying different cryptocurrencies.

%=============================
\section{Discussion} 
\label{Discussion}
%=============================-

We in this section recognise and discuss limitations. 

While relying on time series features is effective, these features may be confounded by traffic obfuscation techniques (e.g., adding proxy, load encryption, port replacement, and packet padding), which may be used by tricky miners. Such techniques could mask cryptomining activities, thereby challenging the accuracy and effectiveness of existing detection schemes. According to our research, cryptomining technology is still in the stage from plaintext transmission to encrypted transmission but has not yet advanced to the stage of encrypted traffic obfuscation.} Most of the mainstream mining software, such as xmrig, CGminer, BFGminer, etc., support SSL/TLS traffic encryption, but none of them support traffic obfuscation, which is unnecessary for the vast majority of miners. Miners use SSL/TLS traffic encryption in order to protect their privacy, while traffic obfuscation is used to avoid regulation. A few 'illegal' miners use privately developed software with these features. We are keeping a close eye on this area and will conduct more in-depth research to evaluate the performance of our approach in traffic obfuscation scenarios.

We also acknowledge that advanced evasion techniques such as encrypted proxy chaining, VPN tunnelling, and protocol mimicry pose additional challenges. These techniques can obscure traffic patterns by layering encryption, rerouting through multiple nodes, or emulating benign protocols, thereby reducing the effectiveness of traditional feature-based classifiers. Although our current system leverages robust time-series features and active probing, it may face limitations in scenarios where traffic is fully encapsulated within VPN tunnels or mimics non-mining application protocols. To address this, future versions of CryptoCatch may incorporate TLS fingerprinting, temporal behavioural modelling, or side-channel features (e.g., packet burst patterns or inter-flow correlation) to detect mining activity hidden under protocol disguise. We consider these directions important next steps in extending CryptoCatch’s robustness for real-world deployment.

For privacy and ethical considerations, our two-step detection mechanism should be deployed at either the main gateway exit of the campus network or at the gateway exit of a specific building. This deployment method will not interfere with existing traffic detection devices and can operate independently. Additionally, the traffic through the gateway exit of the campus network is huge, and our detection mechanism works offline. Our detection mechanism requires only the 5-tuple, timestamp, and payload length of each traffic flow as input, which has minimal impact on the privacy of campus network users. Finally, for the ethical issue, the mining scenario within the campus network is more complex, and our mechanism is unable to distinguish the mining behaviour of researchers and real miners. Thus, further confirmation by other methods is necessary to determine how to deal with these  miners, such as considering the duration of mining and the number of devices accessing the same mining pool in the same place.

Our experiments were conducted primarily in a controlled LAN environment, which provides clean conditions for repeatable measurements and avoids interference from unrelated traffic. However, such an environment may not fully represent real-world network conditions, where background traffic, service diversity, and transient congestion can alter flow characteristics. Although we include supplementary evaluations using mixed campus traffic combined with  CIC-IDS-2017 traffic dataset, we acknowledge that additional measurements in more diverse and dynamic operational networks will be necessary to further assess generalizability in future work.

Mining pools may deploy basic countermeasures to detect or suppress active probing, which could reduce the reliability of probing-based confirmation. Such defenses include token or key validation (responding only to authorized clients), silent dropping of malformed or unknown requests, per-IP connection and rate limiting. While most pools in our study did not actively enforce these measures, it represents a practical limitation that future detection systems must consider.

Future research should aim to address these limitations. This may be achieved by expanding the dataset to include a broader range of cryptomining activities and exploring more robust features that are resistant to obfuscation techniques. Further work is also needed to adapt and test our methodology in real-world network environments, considering the dynamic nature of cryptomining and its implications on network infrastructure.

%================================
\section{Conclusion} 
\label{sec7}
%================================

We introduced CryptoCatch, a two-stage mechanism designed specifically for detecting cryptomining traffic. We first utilized finely-tuned time series features to effectively train XGBoost classifiers, which consistently achieved an F1-score of 0.99 in detecting mining traffic and a 99.39\% accuracy rate across various multi-class cryptocurrency classification scenarios. Then, we developed and assessed an active probing technique aimed at reducing the false positive rate of our classifiers.  Evaluations demonstrate that CryptoCatch effectively identified 98.66\% of mining pool URLs within our dataset.

%==========================================================
\bibliographystyle{unsrt}
\bibliography{bib}
%==========================================================

\end{document}